\documentclass[%
 reprint,
superscriptaddress,
showpacs,preprintnumbers,
 amsmath,amssymb,
 aps,
]{revtex4-1}
\usepackage{soul}
\usepackage{ulem}
\usepackage{graphicx}
\usepackage{dcolumn}
\usepackage{bm}
\usepackage{color}
\newcommand\redout{\bgroup\markoverwith{\textcolor{red}{\rule[.5ex]{2pt}{0.4pt}}}\ULon}

\usepackage[colorlinks=true,citecolor=blue]{hyperref}
\hypersetup{colorlinks=true,citecolor=blue,linkcolor=red,urlcolor=blue}
\usepackage {xcolor}
\begin{document}
\date{\today}
\title{Electrical and thermal transport in a twisted heterostructure of transition metal dichalcogenide and CrI$_3$
connected to a superconductor}
\author{Leyla Majidi}
\email{Leyla.Majidi@ipm.ir}
\affiliation{School of Physics, Institute for Research in Fundamental Sciences (IPM), Tehran 19395-5531, Iran}
\author{Reza Asgari}
\email{asgari@ipm.ir}
\affiliation{School of Physics, Institute for Research in Fundamental Sciences (IPM), Tehran 19395-5531, Iran}
\affiliation{School  of  Physics,  University  of  New  South  Wales,  Kensington,  NSW  2052,  Australia}
\date{\today}

\begin{abstract}
The broad tunability of the proximity exchange effect between transition-metal dichalcogenides (TMDCs) and chromium iodide (CrI$_3$) heterostructures offers intriguing possibilities for the use of TMDCs in two-dimensional magnetoelectrics. In this work, the influence of the twist angle  and the gate electric field on the electric and thermal transport in a TMDC/CrI$_3$ junction is investigated using the Dirac -Bogoliubov-de Gennes equation. We show that significant amounts can be controlled by spin-splitting of band structures due to spin-orbit interaction, and that the exchange-splitting of bands arises from the proximity effect. The property of the Andreev reflection (AR) process is highly dependent on the spin valley polarized states due to spin-orbit coupling. Remarkably, perfect spin valley polarized AR is possible over a wide bias range by using a gate voltage to tune the local Fermi energy and varying the type of charge doping. The proposed structure with $p$-type doping is found to have larger spin valley polarized Andreev conductance and high thermal conductance. We further show that, depending on the TMDC material and chemical potential of the TMDC/CrI$_3$ layer, twisting can lead to suppression or a significant increase in Andreev conductance as well as enhancement of thermal conductance for chemical potentials smaller than that of the superconducting regime.
\end{abstract}
%
\maketitle
\section{Introduction}
The fascinating research on two-dimensional (2D) materials has recently experienced a modern stage in its development, represented by a new degree of freedom in terms of material structure: A twist angle between adjacent layers that facilitates fine-tuning of electronic properties of van der Waals (vdW) heterostructures~\cite{Lopes, Suarez, Luis, Bistritzer}. The most prominent example is the magic-angle twisted bilayer graphene, which exhibits magnetism~\cite{Sharpe} and superconductivity~\cite{Cao,Balents} due to strong correlations. vdW heterostructures, comprising a variety of 2D layered materials, have emerged as potential building blocks for future ultrafast and low-power electronic and spintronic devices~\cite{vdW1,vdW2,vdW3}.

Twistronics is now demonstrating its potential for proximity effects. For 2D spintronics and the emerging field of superconducting spintronics, it is desirable to integrate 2D materials such as graphene and transition metal dichalcogenides (TMDCs) with 2D magnets. It turns out that twisting has a profound influence on both the size and the nature of the proximity-induced spin-orbit coupling (SOC) in the graphene/TMDC heterostructure~\cite{Naimer} as well as on the proximity exchange coupling in graphene/Cr$_2$Ge$ _2$Te$_6$ heterostructure and formation of antiferromagnetic Dirac bands in graphene~\cite{Zollner}.

Semiconducting TMDCs such as MX$_2$ (M=W,Mo;X=S,Se,Te) possess strong SOC and the potential for coupling spin and valley physics arising from it and broken inversion symmetry~\cite{TMDC1,TMDC2,TMDC3,TMDC4}. Recent experiments show a proximity exchange of a few meV in TMDC and CrI$_3$ heterostructures. Interestingly, the CrI$_3$ monolayer is a ferromagnet~\cite{Izadi,Huang,PJiang}, while the CrI$_3$ bilayer exhibits an antiferromagnetic coupling~\cite{Huang,SJiang,Song}. Unlike thin films of conventional ferromagnets, the 2D layered ferromagnets exhibit out-of-plane magnetization, which is a time-reversal breaking analog of the Zeeman splitting in TMDCs. Remarkably, wide tunability of the proximity exchange coupling has been reported in TMDC/CrI$_3$ heterostructures through controlling the interlayer twist between layers and electrostatic gating~\cite{Fabian19}.

On the other hand, the transfer of the low-energy electrons through a nonsuperconducting/superconducting transition is dominated by a peculiar and fascinating type of fundamental reflection process, the Andreev reflection (AR)~\cite{Andreev,de Jong}, in which an incident electron with spin polarization $s$, when hitting the interface, is retro-reflected as a hole with opposite spin polarization $\bar{s}$ and a Cooper pair is transferred to the superconductor. Since the Cooper pair carries a charge of 2e and no heat, the peculiar AR process leads to an increase in electrical conductance and suppression of thermal conductance~\cite{BTK,Riedel,Lee}. In particular, due to the interplay of superconductivity and the unique electronic structure of atomically thin 2D material, novel interesting phenomena appear when superconducting heterostructures are realized in 2D crystals. Many theoretical and experimental works have found peculiar AR processes at low energies~\cite{Beenakker06, Majidi_G1,Majidi_G2,Majidi_TI1,Majidi_TI2, majidi14_1,Efetov,Majidi20} and nonlocal processes have been predicted to occur under different conditions~\cite{Cayssol, Linder09, Majidi_G1, Majidi_G2, Wang, linder14, majidi14_2, Pandey}.

Motivated by the aforementioned physical phenomena, the main aim of this article is to present a comprehensive theoretical study of charge and heat transport through a ferromagnetic (F) /superconducting (S) junction made of proximity exchange-coupled TMDC/CrI$_3$ heterostructure as ferromagnet and superconducting TMDC (with WSe$_2$ and MoSe$_2$ as TMDC) and to focus on the effect of charge doping, twisting the CrI$_3$ monolayer relative to TMDC layer, and applying a transverse electric field to the magnetized TMDC, and finding a suitable experimental set up with desired transport properties. Indeed, these have a wide range of uses. Magnetic-superconducting heterostructures based on 2D vdW materials are the vital building blocks in the design of ultracompact spintronic and superconducting spintronic devices, with potential applications in quantum computing~\cite{vdW1,vdW2,vdW3,Linder2015,Yang,Blamire}. On the other hand, twisting the neighboring layers with respect to each other facilitates the fine tuning of electronic properties of the vdW heterostructures while the individual materials also preserve a great degree of autonomy. Importantly, wide tunability of the proximity exchange in monolayer MoSe$_2$ and WSe$_2$ owing to a ferromagnetic monolayer CrI$_3$ with respect to twisting and gating is an advantage of the TMDC/CrI$_3$ structure relative to other ferromagnetic structures, in which constant and equal proximity-induced exchange interactions are supposed in the conduction and valence bands.

Using Dirac-Bogoliubov-de-Gennes (DBdG) formalism, we reveal the possibility of a perfect AR for a wide range of chemical potentials within the F region by tuning the bias and using the gate voltage to estimate the local Fermi energy and vary the type of charge doping in the F and S regions. For an F/S structure with $p$- or $n$-type doping, a perfect AR can be electrically controlled and visualized over a wide bias range; it can be achieved for subgap biases ($\textit{eV}_{bias}<\Delta_S$) at the defined F-region chemical potential, $\mu_F$, and for a wide range of $\mu_F$, if $\textit{eV}_{bias}=\Delta_S$. In the case of F and S regions with different types of charge doping, a perfect AR can be achieved by raising the chemical potential of the S regions of almost all values of $\mu_F$ if $\textit{eV}_{bias} =\Delta_S$. Due to the existing spin-orbit interaction in the TMDC layer, the AR process with full spin valley polarization is equally possible for a defined range of the F-region chemical potential while considerable damping of the Andreev conductance and hence imperfect AR occurs in the absence of the SOC term in the F region.

We show that the probability of AR and accordingly the Andreev differential conductance can be suppressed in the spin valley polarized AR regime and increased in the nonpolarized regime by twisting. Most importantly, the rotation angle-induced changes in Andreev conductance strongly depend on the value of the subgap bias as well as the TMDC material: It increases with bias at the defined F-region chemical potential in the MoSe$_2$-based structure, while it decreases with prestress and tends to zero for other values of $\mu_F$ in both MoSe$_2$- and WSe$_2$-based structures when $\textit{eV}_{bias} =\Delta_S$. Furthermore, the presence of the positively applied electric field in the ferromagnetic region dampens the Andreev conductance by increasing the proximity exchange coupling.

Next, we examine the heat transport properties of the proposed structure by evaluating the thermal conductance. The thermal conductance maintains an increasing behavior with temperature and disappears at temperatures well below the critical temperature of the superconducting order parameter. Importantly, twisting can lead to a significant increase in thermal conductance in F/S structures with a $p$-doped F region and a decrease in it in the corresponding structure with $n$-type doping because the chemical potentials of the F region are smaller than those of the superconducting region. However, gating slightly reduces the thermal conductance of $p$-doped structures and does not cause significant changes in the case of $n$-doped F/S junctions. Furthermore, we analyze the influence of the type of charge doping on the transport properties of the proposed structure and demonstrate the high charge and thermal conductance of the WSe$_2$ and MoSe$_2$ based structures with $p$-type doping.

Finally, we compare the results of the WSe$_2$-based structure with those of the corresponding MoSe$_2$-based structure and observe the enhancement of the thermal conductance and damping of the Andreev conductance in the $p$-doped MoSe$_2$-based structure as well as reducing the thermal conductance with the $n$-doped structure. We should note that different qualitative behaviors arise for the chemical potential dependence of both Andreev and thermal conductance by varying the type of charge doping in the F and S regions and the SOC term in the F region. In addition, changing the TMDC material also leads to a different $\mu_F$-dependence of the thermal conductance.

The rest of the paper is arranged as follows. Section \ref{sec:II} is devoted to the theoretical model and fundamental formalisms which will be implemented to investigate AR in a twisted TMDC/CrI$_3$-superconducting TMDC junction. In Sec. \ref{sec:III}, we present our numerical results for the Andreev differential conductance and the thermal conductance of the proposed structures. Finally, a brief summary of results is given in Sec. \ref{sec:IV}.

\section{Theoretical formalism}\label{sec:II}
We consider a wide ballistic F/S junction normal to the $x$ axis with proximity exchange-coupled TMDC/CrI$_3$ heterostructure (as F region) for $x<0$ and superconducting TMDC (as S region) extending over the $x>0$ region (with WSe$_2$ and MoSe$_2$ as TMDC). The magnetic insulator substrate CrI$_3$ is weakly coupled to the TMDC by vdW forces, preserving the characteristic electronic band structure of the TMDC. The proximity exchange coupling splits the conduction and valence bands of the TMDC by roughly 1-5 meV and combined with the intrinsic (valley Zeeman) SOC of the TMDC lifts the valley degeneracy~\cite {Fabian19}. The effective low-energy Hamiltonian which describes the proximity effects in TMDC due to CrI$_3$, has the form~\cite{Fabian19}\begin{eqnarray}
\label{H}
&&\mathcal{H}=\mathcal{H}_0+\mathcal{H}_{\Delta}+\mathcal{H}_{soc}+\mathcal{H}_{ex}+\mathcal{H}_R,\\
&&\mathcal{H}_0=\hbar v_{\rm F}s_0\otimes(\tau \sigma_x k_x+\sigma_y k_y),\\
&&\mathcal{H}_{\Delta}=\frac{\Delta}{2}s_0\otimes\sigma_{z},\\
&&\mathcal{H}_{soc}=\tau s_z\otimes(\lambda_c\sigma_{+}+\lambda_v\sigma_{-}),\\
&&\mathcal{H}_{ex}=-s_z\otimes(B_c\sigma_{+}+B_v\sigma_{-}),\\
&&\mathcal{H}_R=\lambda_R(\tau s_y\otimes\sigma_x-s_x\otimes\sigma_y).
\end{eqnarray}
Here $\tau=\pm1$ is the valley index for the K (K') point and $v_{\rm F}$ represents the Fermi velocity. The pseudospin Pauli matrices $\sigma_i$ ($i = 0,x,y,z$) act on the conduction and valence-band subspaces and $s_i$ ($i = 0,x,y,z$) refers to real spin. The parameter $\Delta$ denotes the orbital gap of the spectrum. The spin-splitting of the conduction and valence bands due to the intrinsic SOC are determined by the parameters $\lambda_c$ and $\lambda_v$, respectively. Proximity exchange effects are presented with $B_c$ and $B_v$ describing the proximity-induced exchange splitting of the conduction and valence bands. The Rashba SOC parameter $\lambda_R$ is due to the presence of the inversion asymmetry in the heterostructure. For the sake of simplicity, we introduce $\sigma_{\pm} = (\sigma_0 \pm \sigma_z)/2$.
\par
Recently, wide tunability of the proximity exchange in monolayer MoSe$_2$ and WSe$_2$ owing to a ferromagnetic monolayer CrI$_3$ has been explored with respect to twisting and gating~\cite{Fabian19}. In particular, proximity exchange splittings depend on the twist angle between the TMDC and the CrI$_3$. Not only do the magnitudes of the exchange differ, but also, remarkably, the direction of the exchange field for the valence band changes sign. Moreover, the proximity exchange parameters increase when the electric field is turned from negative to positive values, which enables the gate control of proximity exchange. We should mention that the gap parameter as well as the Fermi velocity are unaffected by external electric fields.
\par
The S part can be produced by depositing a superconducting electrode on the top of the TMDC sheet~\cite{Ye}. In this region, exchange parameters are zero and the proximity-induced superconducting correlations are characterized by the superconducting pair potential (order parameter) $\Delta_S$.
\par
We assume perfectly flat and clean interfaces in the proposed structure. Transport takes place in the $x$ direction and is described by the DBdG equation~\cite{Beenakker06,majidi14_1}
\begin{eqnarray}
\label{DBdG}
\hspace{-5mm}\Bigg(
\begin{array}{cc}
\mathcal{H}+U(\bm{r})-\mu_0 & \tilde{\Delta}_S(\bm{r})\\
{\tilde{\Delta}_{S}}^{\dag}(\bm{r})& \mu_0-[\mathcal{T}\mathcal{H}\mathcal{T}^{-1}+U(\bm{r})]
\\
\end{array}
\Bigg)
\Bigg(
\begin{array}{c}
u\\
v
\end{array}
\Bigg)
=\varepsilon\Bigg(
\begin{array}{c}
u\\
v
\end{array}
\Bigg),\nonumber\\
\end{eqnarray}
where $\varepsilon$ is the excitation energy, $\mu_0$ is the chemical potential, $U(\bm{r})$ denotes an electrostatic potential and $\mathcal{T}$ represents the time-reversal operator. Substituting the effective single-particle Hamiltonian $\mathcal{H}$ [Eq. ({\ref{H}})], time-reversal operator $\mathcal{T}=i\tau_x \sigma_y \mathcal{K}$ (with $\mathcal{K}$ the operator of the complex conjugation, and $\tau_x$ the Pauli matrix acting on the valley space), and the pair potential $\tilde{\Delta}_S(\bm{r})= \Delta_S(\bm{r})\ e^{i\varphi} s_0\otimes\sigma_x$ into Eq. ({\ref{DBdG}}) results in two decoupled sets of 4D DBdG equations for given spin $s$ and valley $\tau$, which each possesses the form~\cite{majidi14_2}
\begin{equation}\label{DBDG2}
\hspace{-0.5cm}\Bigg(
\begin{array}{cc}
H_{s,\tau}^e-\mu(\bm{r}) & \Delta_S(\bm{r})\ e^{i\varphi}\\
\Delta_{S}(\bm{r})\ e^{-i\varphi}& \mu(\bm{r})-H_{s,\tau}^h
\\
\end{array}
\Bigg)
\Bigg(
\begin{array}{c}
u_{s,\tau}\\
v_{\bar{s},\bar{\tau}}
\end{array}
\Bigg)
=\varepsilon\Bigg(
\begin{array}{c}
u_{s,\tau}\\
v_{\bar{s},\bar{\tau}}
\end{array}
\Bigg),
\end{equation}
where $\mu(\bm{r})=\mu_0-U(\bm{r})$. Considering the electrostatic potential $U(\bm{r})$ to be $-U_0$ in the S region and $U(\bm{r}) = 0$ in the F region, the doping profile throughout the junction is set by different but constant chemical potentials in the S and F regions. The same assumption is used in the self-consistent solution of the DBdG equation to avoid a significant amount of complexity to the problem~\cite{Black-Schaffer2008,Halterman2011,Halterman2013}. The electron and hole wave functions, $u_{s,\tau}$ and $v_{\bar{s},\bar{\tau}}$, are two-component spinors of the form $(\psi_c,\psi_v)$ where $c$ and $v$ respectively denote the conduction and valence bands, $\bar{s}=-s$ and $\bar{\tau}=-\tau$. Therefore the electron excitations in one valley are coupled by the superconducting pair potential $\Delta_S$ to hole excitations in the other valley. The superconducting phase $\varphi$ only plays a role in junctions with several superconductors, and hence we omit it here.

The pair potential is zero in the F region, which represents a reasonable approximation if the superconducting coherence length is appreciably larger than the Fermi wavelength. Inside the S region, the proximity induced pair potential $\Delta_S(\bm{r})$ reaches its bulk value $\Delta_S$ at a distance from the interface which becomes negligibly small if the Fermi wavelength in the S region is significantly smaller than the value in the F region. Therefore neglecting the decay of the order parameter in the vicinity of the interface, the step-function model for $\Delta_S(\bm{r})$ is valid for length scales much longer than the superconducting coherence length~\cite{Beenakker06,Beenakker08,Linder2007}. In the case of moderate Fermi level mismatch, self-consistent calculations have demonstrated that increasing the chemical potential of the normal (N) region leads to weekly doping dependence of the local density of states and therefore the pair potential inside the S region of the graphene-based S/N/S structures~\cite{Halterman2011} and confirmed that the sharp interface between the regions can represent an appropriate approximation. Moreover, the self-consistent DBdG approach to accurately determine the spatial profile of the pair potential is favorable in finite sized structures where quantum scale oscillations play a role and has significant consequences for quasiparticle bound states, and supercurrent flow~\cite{Black-Schaffer2008,Halterman2011,Halterman2013}. On the other hand, the doping level as well as the width of the S region can have substantial effects on the critical temperature of the superconductivity, $T_C$. It has been demonstrated by self-consistent calculations that decreasing the Fermi level mismatch, by tuning the doping level in the N region of finite sized graphene-based S/N/S junction, leads to substantial variations of $T_C$ for S regions that occupy a fraction of the coherence length whereas it has no detrimental effect for large lengths of the S region~\cite{Halterman2011}. Besides, the experimental results of the gate-modulated Andreev conductance across the low-disorder vdW interface between graphene and the superconducting NbSe$_2$ are in good qualitative and quantitative agreement with the theoretical model based on the non-self-consistent DBdG equation with step function model for the pair potential and constant chemical potential~\cite{Efetov}. Therefore in the proposed F/S structure with semi-infinite S and F regions, it is reasonable to assume that the pair potential is constant in the S region and zero in the F region and adopts the step-function model of $\Delta_S(\bm{r})=\Delta_S \Theta(x)$ for the pair potential.
\par
To study AR at the F/S interface within the scattering formalism, we first construct the quasiparticle wave functions that participate in the scattering processes. According to Eq. (\ref{DBDG2}), the dynamics of the low-energy itinerant charge carriers inside the F region with spin $s$ from valley $\tau$ are described by effective Hamiltonians $H_{s,\tau}^{e}$ and $H_{s,\tau}^{h}$, respectively, for the electron and hole subsectors,
\begin{eqnarray}
H_{s,\tau}^{e(h)}&=&\hbar v_{\rm F}(\tau \sigma_x k_x+\sigma_y k_y)+\frac{\Delta}{2}\sigma_{z}+ s \tau\ (\lambda_c\sigma_{+}+\lambda_v\sigma_{-})\nonumber\\
&-&s(\bar{s})(B_c\sigma_{+}+B_v\sigma_{-}).
\end{eqnarray}
The Rashba parameter $\lambda_R$ is set to zero since there is no in-plane component of the spin expectation value around the band edges, and therefore it is unnecessary to capture the essentials of the band structure for the TMDC/CrI$_3$ stacks. At a given energy $\varepsilon$ and a transverse momentum $k_y$, the solutions are two states of the form
\begin{equation}
\label{psie}
\psi_c^{e\pm}={A_{s,\tau}^e}\ e^{\pm i\tau k_x^e x} e^{ik_y y}
\left(
\begin{array}{c}
\pm \tau a_{s,\tau}^e e^{\mp i\tau\theta_{s,\tau}^e}\\
1\\
0\\
0
\end{array}
\right),
\end{equation}
for the conduction band electrons with energy-momentum relation
\begin{eqnarray}
\label{Ee}
\varepsilon_c^e&=&-\mu_F+\frac{1}{2}[-s(B_c+B_v)+s \tau(\lambda_c+\lambda_v)\nonumber\\
&+&\sqrt{4 (\hbar v_{\rm F} |\bm{k}_{s,\tau}^e|)^2+([s(B_c+B_v)+\Delta]+s\tau(\lambda_c-\lambda_v))^2}]\nonumber\\
\end{eqnarray}
and
\begin{equation}
\label{psih}
\psi_c^{h\pm}={A_{s,\tau}^h}\ e^{\mp i\tau k_x^h x} e^{ik_y y}
\left(
\begin{array}{c}
0\\
0\\
\mp \tau a_{s,\tau}^h e^{\pm i\tau\theta_{s,\tau}^h}\\
1
\end{array}
\right),
\end{equation}
for the conduction band holes of the $n$-doped F region with energy-momentum relation
\begin{eqnarray}
\label{Eh}
\varepsilon_c^h&=&\mu_F-\frac{1}{2}[-\bar{s}(B_c+B_v)+s \tau(\lambda_c+\lambda_v)\nonumber\\
&+&\sqrt{4 (\hbar v_{\rm F} |\bm{k}_{s,\tau}^h|)^2+([\bar{s}(B_c+B_v)+\Delta]+s\tau(\lambda_c-\lambda_v))^2}].\nonumber\\
\end{eqnarray}
As seen here, the spin-splitting and exchange-splitting of the conduction and valence bands appear in the eigenvalues and eigenvectors and subsequently the transport properties vary with changing those quantities. In Eqs. (\ref{psie}) and (\ref{psih}), $\mu_F=\mu_0$, $a_{s,\tau}^{e(h)}=[\mu_F+(-)\varepsilon+s(\bar{s}) B_v-s\tau\lambda_v+\Delta/2]/\hbar v_{\rm F} |\bm{k}_{s,\tau}^{e(h)}|$, $A_{s,\tau}^{e(h)}=1/\sqrt{2\  a_{s,\tau}^{e(h)}\cos{(\tau \theta_{s,\tau}^{e(h)})}}$, and $\theta_{s,\tau}^{e(h)}=\arcsin({k_y/|\bm{k}_{s,\tau}^{e(h)}|})$ is the angle
of the propagation of the electron (hole). The two propagation directions of the electron (hole) along the $x$-axis are denoted by $\pm$ in $\psi_c^{e(h)\pm}$.
\par
In pristine TMDC monolayers, the two valleys are the same due to the time-reversal symmetry. The vicinity to the 2D magnetic material CrI$_3$ provides a direct and effective way to break the valley degeneracy of TMDC because of the magnetic proximity effect. The presence of negative proximity exchange field in TMDC/CrI$_3$ heterostucture~\cite{Fabian19} shifts the spin-s subband of the $\tau$ ($\bar{\tau}$) valley upward and the spin-$\bar{s}$ subband of the $\tau$ ($\bar{\tau}$) valley downward in the conduction and valence bands, respectively, by $|B_c|$ and $|B_v|$. In contrast, twisting the CrI$_3$ layer with respect to the monolayer TMDC replaces the sign of the proximity exchange in the valence band (see Table. \ref{table1}) and therefore brings the spin-subbands of the $\tau$-valley close to each other and gets those of the $\bar{\tau}$ valley far away from each other. Therefore the proximity exchange coupling adjusts the valley splitting and the TMDC/CrI$_3$ heterostructure shows valley polarization of the TMDC~\cite{majidi14_2,Majidi_MOS2_3}.
\par
Inside the S region, the solutions are rather mixed electron-hole excitations (called Dirac-Bogoliubov quasiparticles) that either decay exponentially as
$x \rightarrow\infty$ (for subgap solutions when $\varepsilon \leq \Delta_S$) or propagate along the $x$ direction (for supragap solutions when $\varepsilon>\Delta_S$). These solutions for the $n$-doped S region take the form
\begin{equation}
\psi_c^{S\pm}=e^{ i\tau k'_{\pm, x}x} e^{ik_y y}
\left(
\begin{array}{c}
  u_{1\pm} \\
  u_2 \\
  u_{3\pm}\\
 1
\end{array}
\right),
\end{equation}
where
\begin{eqnarray}
u_{1\pm}&=&\frac{1}{4\hbar v_{\rm F} \Delta_S (\tau k'_{\pm, x}+ik_y)}[4(\hbar v_{\rm F} k_{S\pm})^2+\Delta^2\nonumber\\
&-&\frac{1}{2\mu_S-s\tau(\lambda_c+\lambda_v)}[s\tau(\lambda_c-\lambda_v)[-2\mu_S+s\tau(\lambda_c+\lambda_v)]\nonumber\\
&-&B](-2\mu_S+2s\tau\lambda_c+2\varepsilon)+\Delta(-2\mu_S+s\tau(3\lambda_c-\lambda_v)\nonumber\\
&+&\frac{B}{2\mu_S-s\tau(\lambda_c+\lambda_v)}+2\varepsilon)],\\
u_2&=&\frac{1}{4\mu_S\Delta_S-2s\tau\Delta_S(\lambda_c+\lambda_v)}(B+[2\mu_S-s\tau(\lambda_c+\lambda_v)]\nonumber\\
&\times&[-2\mu_S+s\tau(\lambda_c+\lambda_v)+2\varepsilon]),\\
u_{3\pm}&=&-\frac{1}{2\hbar v_{\rm F}(\tau k'_{\pm, x}+ik_y)[-2\mu_S+s\tau(\lambda_c+\lambda_v)]}([\Delta\nonumber\\
&+&s\tau(\lambda_c-\lambda_v)][2\mu_S-s\tau(\lambda_c+\lambda_v)]+B),
\end{eqnarray}
$\mu_S=\mu_0+U_0$, $k'_{\pm, x}=\pm k_0+i\kappa \tau$, $k_{S\pm}={(2\hbar v_{\rm F})}^{-1}[4 (\mu_S^2+\varepsilon^2-\Delta_S^2-s\tau\mu_S(\lambda_c+\lambda_v)+\lambda_c\lambda_v\pm\sqrt{(\varepsilon^2-\Delta_S^2)[2\mu_S-s\tau(\lambda_c+\lambda_v)]^2})+
2s\tau\Delta(\lambda_v-\lambda_c)-\Delta^2$, and $B=([-2\mu_S+s\tau(\lambda_c+\lambda_v)]^2[4(\hbar v_{\rm F} k_{S\pm})^2+\Delta^2+2s\tau\Delta(\lambda_c-\lambda_v)+(\lambda_c-\lambda_v)^2])^{1/2}$.
\par
An incident electron from the conduction band of $n$-doped F region with a subgap energy $0 \leq \varepsilon \leq \Delta_S$ could undergo two possible scattering events. It can either be normally reflected as an electron in the conduction band via normal reflection (NR) or be Andreev reflected as a hole in the same band with opposite spin and different valley index via retro AR. Depending on the magnitude of the chemical potential $\mu_F$ and the excitation energy $\varepsilon$, the incident electron and the reflected hole can be from one or two of the spin subbands. As long as $E_c^{s,\bar{\tau}}-\varepsilon\leq\mu_F<E_c^{s,\tau}-\varepsilon$ (with $E_c^{s,\bar{\tau}}=B_c+\lambda_c+\Delta/2$ and $E_c^{s,\tau}=-B_c+\lambda_c+\Delta/2$, respectively, the energies of the conduction band edges for spin-$s$ subbands of the $\bar{\tau}$ and $\tau$ valleys), only the lower spin subbands with $s=-\tau=1$ and $s=-\tau=-1$ contribute to the transport of charge and result in a spin valley polarized AR process with $s\tau=-1$. For the case of $\mu_F\geq E_c^{s,\tau}-\varepsilon$, the Fermi level crosses the two spin subbands of $\tau$ and $\bar{\tau}$ valleys with $s\tau=\pm 1$ and, therefore, the AR process is not spin valley polarized while in the case of $p$-type doped F region, the charge transport is determined by incoming electrons from the upper spin subbands of the valence band with $s=\tau=1$ and $s=\tau=-1$, when $E_v^{\bar{s},{\tau}}-\varepsilon<\mu_F\leq E_v^{\bar{s},\bar{\tau}}-\varepsilon$ (with $E_v^{\bar{s},{\tau}}=B_v-\lambda_v-\Delta/2$ and $E_v^{\bar{s},\bar{\tau}}=B_v+\lambda_v-\Delta/2$, respectively, the energies of the valence-band edges for spin $\bar{s}$ subbands of the $\bar{\tau}$ and $\tau$ valleys). Therefore the AR process will be fully spin valley polarized with $s\tau=1$. We should note that the twist-angle induced sign change of the proximity exchange $B_v$ will make the AR process to be spin valley polarized (with $s\tau=1$) for the chemical potentials in the range $E_v^{{s},\bar{\tau}}-\varepsilon<\mu_F\leq  E_v^{{s},{\tau}}-\varepsilon$ (with $E_v^{{s},\bar{\tau}}=-B_v-\lambda_v-\Delta/2$ and $E_v^{{s},{\tau}}=-B_v+\lambda_v-\Delta/2$).

Denoting the amplitudes of the NR and AR processes $r_e^{s,\tau}$ and $r_h^{\bar{s},\bar{\tau}}$, respectively, the total wave functions inside the F and S regions can be written as
\begin{eqnarray}
\label{N wave function}
&&\psi_{F}=\psi_c^{e+}+r_{s,\tau}^e\ \psi_c^{e-}+r_{\bar{s},\bar{\tau}}^h\ \psi_c^{h-},\\
\label{s wave function}
&&\psi_{S}=t\ \psi_c^{S+}+t'\ \psi_c^{S-}.
\end{eqnarray}
Matching the wave functions of the F and S regions at the interface $x =0$, the scattering coefficients for the normal and AR processes can be obtained. Having known the reflection coefficients, we investigate the charge and thermal conductance of the F/S interface in the following section.
\par
We should mention that earlier works in this field~\cite{majidi14_1,majidi14_2,Majidi_MOS2_3}, have studied the transport characteristics of the MoS$_2$-based heterostructures by supposing an equal proximity exchange splitting in the conduction and valence bands, $B_c$ and $B_v$, and ignoring the SOC-induced spin-splitting in the conduction band, $\lambda_c$, while in this paper, we employ an effective Hamiltonian which fully describes the band structure of the TMDC/CrI$_3$ heterostructure (with WSe$_2$ and MoSe$_2$ as TMDC) in the presence of different $\lambda_c$ and $\lambda_v$ as well as different proximity-induced exchange interactions $B_c$ and $B_v$. Significantly, another advantage of the proposed structure is that twisting the monolayer CrI$_3$ relative to the TMDC layer and applying a gate electric field to the TMDC/CrI$_3$ heterostructure are efficient tenable knobs to tailor the sign and magnitude of the proximity exchange interactions.
\begin{table*}[t]
\begin{center}
\begin{tabular}{l|c|c|c|c|r}
\hline\hline
\vline ~ & & & & & \vline\\
\vline ~ \textsl{Structure} & \textsl{$\Delta$(eV)} & \textsl{$\lambda_c$(meV)} & \textsl{$\lambda_v$(meV)} & \textsl{$B_c$(meV)} & \textsl{$B_v$(meV)} \vline \\
\vline ~ & & & & & \vline\\
\hline
\vline ~ & & & & & \vline\\
\vline ~ Bare WSe$_2$& 1.327 &13.90 & 241.79 &0&0   \vline\\
\vline ~ & & & & & \vline\\
\hline
\vline ~ & & & & & \vline\\
\vline ~ WSe$_2$/CrI$_3$ & 1.358 &-&-& -2.223 & -1.446  \vline\\
\vline ~ (no SOC, no twist)& & & & & \vline\\
\hline
\vline ~ & & & & & \vline\\
\vline ~ WSe$_2$/CrI$_3$ & 1.327&13.81&240.99&-1.783&-1.583 \vline\\
\vline ~ (with SOC, no twist)& & & & & \vline\\
\hline
\vline ~ & & & & & \vline\\
\vline ~ WSe$_2$/CrI$_3$ & 1.417  & -& - & -1.648&1.896 \vline\\
\vline ~ (no SOC, with twist $30^\circ$)& & & & & \vline\\
\hline
\vline ~ & & & & & \vline\\
\vline ~ Bare MoSe$_2$&1.302&-9.647 &94.56 &-&-  \vline\\
\vline ~ & & & & & \vline\\
\hline
\vline ~ & & & & & \vline\\
\vline ~ MoSe$_2$/CrI$_3$ & 1.305 &-&-& -2.081 & -1.454  \vline\\
\vline ~ (no SOC, no twist)& & & & & \vline\\
\hline
\vline ~ & & & & & \vline\\
\vline ~ MoSe$_2$/CrI$_3$ & 1.301& -9.678 & 94.43 & -1.592&-1.426 \vline\\
\vline ~ (with SOC, no twist) & & & & & \vline\\
\vline ~ & & & & & \vline\\
\hline
\vline ~ & & & & & \vline\\
\vline ~ MoSe$_2$/CrI$_3$ & 1.351  & -& - & -1.641& 0.502 \vline\\
\vline ~ (no SOC, with twist $30^\circ$)& & & & & \vline\\
\hline\hline
\end{tabular}
\caption{The orbital gap $\Delta$, SOC $\lambda_{c(v)}$ and the proximity exchange $B_{c(v)}$ of the conduction (valence) band for bare TMDC and TMDC/CrI$_3$ heterostructure (with WSe$_2$ and MoSe$_2$ as TMDC) in the presence or absence of the twist and SOC~\cite{Fabian19}. Notice that $B_{v}$ changes sign in twisted cases.}
\label{table1}
\end{center}
\end{table*}
\section{Numerical results and discussion}\label{sec:III}
In this section, we mainly study the charge and thermal transport in the proposed F/S interface with the subgap energy regime. We concentrate on the Andreev differential conductance as well as the thermal conductance in the junction. Before presenting our numerical results, we should mention that the orbital gap $\Delta$, SOC $\lambda_{c(v)}$, and the exchange interaction $B_{c(v)}$ parameters for different TMDCs (like WSe$_2$ and MoSe$_2$) are set to the values obtained in Ref. ~\cite{Fabian19} (see Table. \ref{table1}).

Due to the vicinity to the 2D magnetic material, the magnetization direction of the TMDC is the same as in the I atoms of the monolayer CrI$_3$, but opposite to the Cr atoms, resulting in a negative proximity exchange parameter. The absence or presence of the SOC term in the heterostructure calculations does not affect significant changes in the magnitude of the exchange parameters. However, twisting the CrI$_3$ layer relative to the TMDC changes the sign of the exchange field in the valence band and makes the valence-band spin-splitting to be opposite in sign in the absence of SOC. The magnitude of the proximity exchange $B_v$ increases with twisting in the WSe$_2$-based heterostructure whereas it decreases in the MoSe$_2$-based heterostructure. Therefore twisting can remain an effective tool to modify the proximity exchange field. Since the parameters in the presence of the SOC in the heterostructure are barely varied from those of the heterostructure without SOC, we set $\Delta=1.417 (1.351)$ eV, $\lambda_{c}=13.81 (-9.675)$ meV, $\lambda_{v}=240.99 (94.43)$ meV, $B_c=-1.648 (-1.641)$ meV and $B_v=1.896 (0.502)$ meV for the twisted WSe$_2$ (MoSe$_2$)/CrI$_3$ heterostructure with SOC.
\par
The chemical potentials $\mu_F$ and $\mu_S$ are in units of electron volt (eV). For practical applications in electronic devices, the single layer and multilayer TMDCs can be $n$- or $p$-type doped on generating desirable charge carriers~\cite{Radisavljevic,Fontana}. We set the zero-temperature superconducting order-parameter $\Delta_S = 1$ meV and scale the excitation energy, $\varepsilon$, in units of $\Delta_S$ and the temperature $T$ is in units of the critical temperature of the superconducting order-parameter $T_C$.
\subsection{Electrical conductance}
\begin{figure}[]
\begin{center}
\includegraphics[width=3.4in]{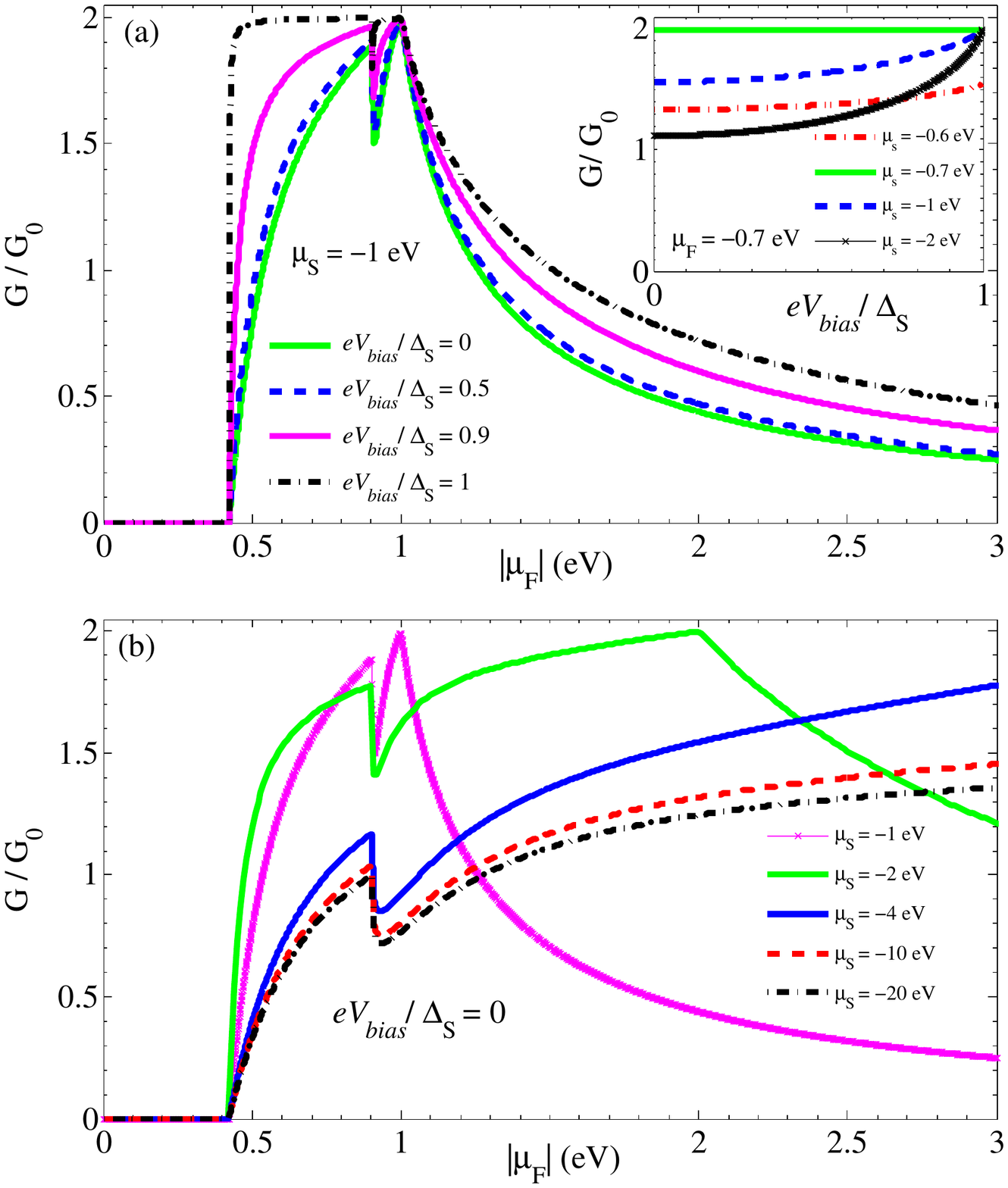}
\end{center}
\caption{\label{Fig:1} (Color online) (a) The normalized Andreev differential conductance of the WSe$_2$-based structure with $p$-doped F and S regions versus the magnitude of the chemical potential inside the F region, $|\mu_F|$, for different values of the subgap bias $\textit{eV}_{bias}/\Delta_S$, when the chemical potential of the S region is set to $\mu_S = -1$ eV. Inset of (a) presents the behavior of the Andreev conductance versus $\textit{eV}_{bias}/\Delta_S$ for different values of $\mu_S$, when $\mu_F=-0.7$  eV. (b) The behavior of the Andreev conductance versus $|\mu_F|$ for different values of the chemical potential $\mu_S$, when $\textit{eV}_{bias}/\Delta_S = 0$. Note that there is no twist between the CrI$_3$ and the WSe$_2$ in the proposed structure.}
\end{figure}
\begin{figure}[]
\begin{center}
\includegraphics[width=3.5in]{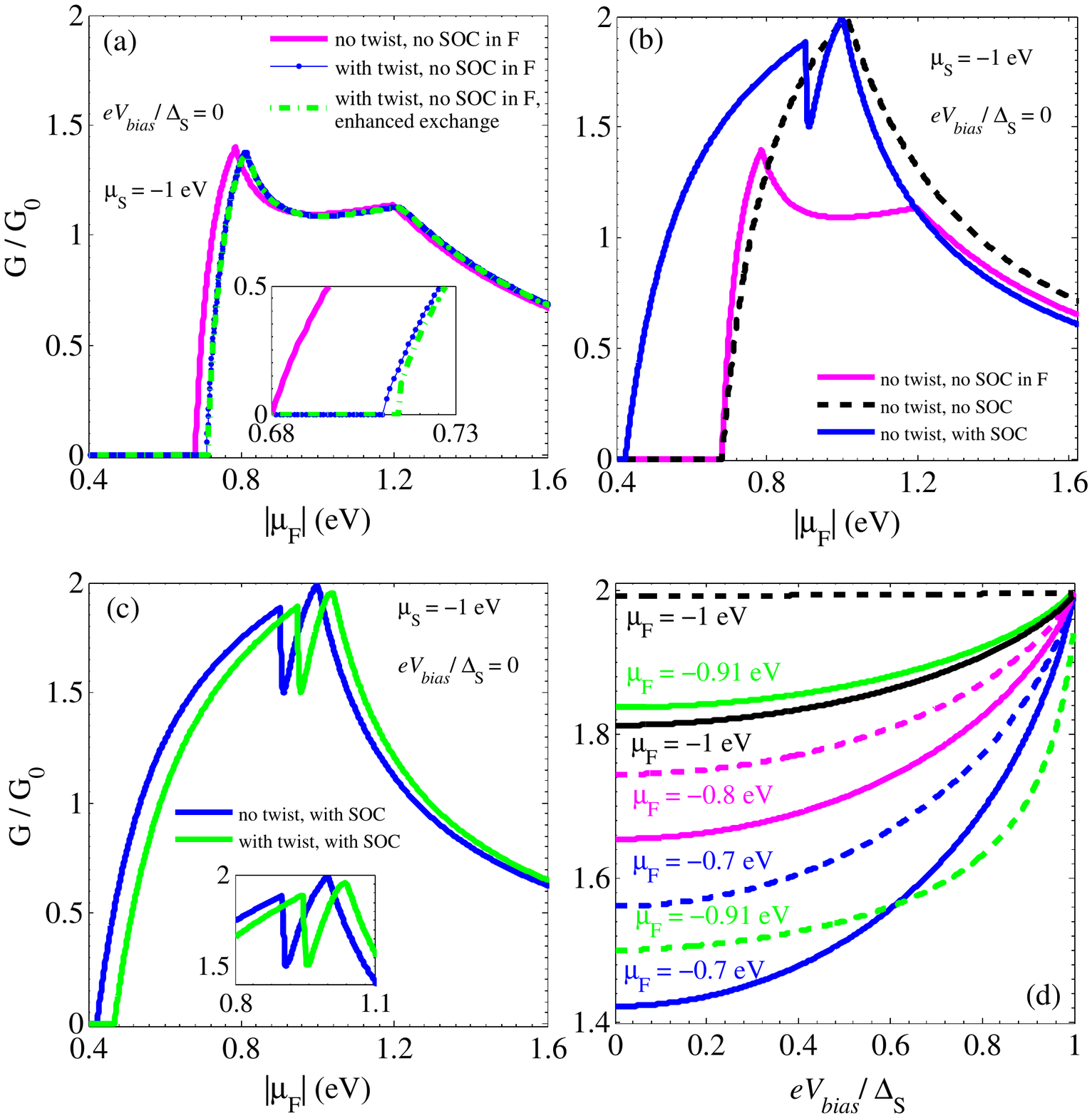}
\end{center}
\caption{\label{Fig:2} (Color online) (a) The effects of twisting and the applied transverse-electric field on the chemical potential $|\mu_F|$ dependence of the zero-bias Andreev conductance in the absence of the SOC inside the F region of the WSe$_2$-based $p$-doped F/S structure, when $\mu_S=-1$ eV. The twist angle between the CrI$_3$ and the WSe$_2$ is $30^{\circ}$. The enhancement of the magnitude of the proximity exchange interaction in the presence of the positive transverse-electric field is about $4$ meV. The effect of the SOC on the zero-bias Andreev conductance in the (b) absence and (c) presence of twisting. (d) The bias voltage dependence of the Andreev conductance for the $0^{\circ}$ (dashed lines) and $30^{\circ}$ (solid lines) twist angles between the CrI$_3$ and the WSe$_2$ in WSe$_2$/CrI$_3$ heterostructure, when $\mu_F=-0.7, -0.8, -0.91$ and $-1$ eV, and $\mu_S = -1$ eV.}
\end{figure}
To study electrical transport, we assume that the F region is in contact with a biased reservoir, and that the S region is in contact with a reference reservoir. Applying a bias voltage $\textit{V}_{bias}$ through the junction induces an electric current. In Blonder-Tinkham-Klapwijk formalism~\cite{BTK}, the Andreev differential conductance at zero temperature reads as
\begin{eqnarray}
\label{G}
\hspace{-2mm}G(\textit{V}_{bias})&=&\sum_{s,\tau=\pm1}G_{0}^{s,\tau}(\textit{V}_{bias})\int_{0}^{\theta_{s,\tau}^c}[1-|r_{s,\tau}^e(\theta_{s,\tau}^e,\textit{eV}_{bias})|^2\nonumber\\
&+&|r_{\bar{s},\bar{\tau}}^h(\theta_{s,\tau}^e,\textit{eV}_{bias})|^2]\cos{\theta_{s,\tau}^e}\ d\theta_{s,\tau}^e,
\end{eqnarray}
where $G_{0}^{s,\tau}(\textit{V}_{bias})={e^2 W|\bm{k}_{s,\tau}^{e(h)}(\textit{eV}_{bias})|}/{\pi h}$ characterizes the spin-$s$ valley-$\tau$ conductance of a TMDC/CrI$_3$ heterostructure of width $W$ with perfect transmission. Here $\theta_{s,\tau}^{c}=\arcsin({|\bm{k}_{s,\tau}^h|}/{|\bm{k}_{s,\tau}^e|})$ is the critical angle of incidence above which the Andreev reflected waves become evanescent and do not contribute to any transport of charge. Herein, we have put $\varepsilon=\textit{eV}_{bias}$ at zero temperature. Also, it is convenient to introduce the normalized conductance $G/G_0$ with $G_0 =\sum_{s,\tau=\pm1}G_{0}^{s,\tau}$.
\par
Figure \ref{Fig:1}(a) shows the behavior of the normalized Andreev conductance $G/G_0$ of $p$-doped WSe$_2$-based structure in terms of the magnitude of the chemical potential inside the F region $|\mu_F|$ when the chemical potential of the S region is set to $\mu_S = -1$ eV. There is a wide gap in Andreev conductance owing to the orbital gap in the band structure of WSe$_2$. Increasing the chemical potential of the F region leads to an enhancement of the zero-bias Andreev conductance and perfect AR with $G/G_0=2$ occurs at $|\mu_F|=|\mu_S|$. Henceforth, it undergoes a decreasing behavior for $|\mu_F|>|\mu_S|$. Significantly, a cusplike behavior occurs at the edge of the spin-$\bar{s}$ valley-$\tau$ of the valence band with $E_v^{\bar{s},\tau}=B_v-\lambda_v-\Delta/2$, such that for $|\mu_F|<|E_v^{\bar{s},\tau}-\textit{eV}_{bias}|$ the AR process is spin valley polarized (spin-polarized in each valley, with $s\tau=1$). For subgap biases, the probability of the AR process and, therefore, the Andreev conductance increase by enhancing the $\textit{eV}_{bias}/\Delta_S$ ratio and perfect AR occurs for a broad range of the chemical potential $|\mu_F|\leq|\mu_S|$, when $\textit{eV}_{bias}/\Delta_S = 1$. The inset of Fig. \ref{Fig:1}(a) demonstrates an enhancement of the Andreev conductance with respect to the $\textit{eV}_{bias}/\Delta_S$ ratio for different values of the chemical potential $\mu_S$, such that perfect AR happens for all values of the subgap biases when $|\mu_S|=|\mu_F|$. The Andreev conductance can be enhanced or reduced by increasing the chemical potential of the S region depending on the value of the bias voltage $\textit{eV}_{bias}/\Delta_S$ ratio. Moreover, it is presented in Fig. \ref{Fig:1}(b) that the behavior of the Andreev conductance versus $|\mu_F|$ is extremely sensitive to the value of the chemical potential inside the S region. The Andreev conductance has an increasing behavior with $|\mu_F|$ for large values of $\mu_S$, while it experiences a decreasing behavior after a sharp peak at $|\mu_F|=|\mu_S|$ for smaller values of $\mu_S$.
\par
In the following, we evaluate the effect of the twist angle between the CrI$_3$ and the WSe$_2$ as well as the applied transverse-electric field on the AR process and accordingly the Andreev conductance in the proposed structure. To obtain the effect of the twist angle, we first turn off the SOC parameter [see Fig. \ref{Fig:2}(a)]. The absence of the SOC in the F region (magnetized gapped graphene/superconducting WSe$_2$ structure) leads to various features of the Andreev conductance; it attenuates the Andreev conductance considerably, prevents the perfect AR process (with $G/G_0=2$), and removes the cusp-like behavior of the Andreev conductance. Most importantly, the twist angle-induced sign change of the proximity exchange in the valence band, $B_v$, leads to the suppression of the AR process and accordingly the Andreev conductance immediately after the wide gap in comparison to that of the untwisted structure. In particular, enhancement of the magnitude of the proximity exchange interaction by applying a positive gate electric field reduces the Andreev conductance and suppresses it in a broader range of the chemical potential $\mu_F$ than that of the twisted one [see the inset of Fig. \ref{Fig:2}(a)]. In addition, it can be noted from Fig. \ref{Fig:2}(b) that the Andreev peak reappears at $|\mu_F|=|\mu_S|$ if we switch off the SOC term in both F and S regions (gapped graphene-based structure).
\par
\begin{figure}[]
\begin{center}
\includegraphics[width=3.4in]{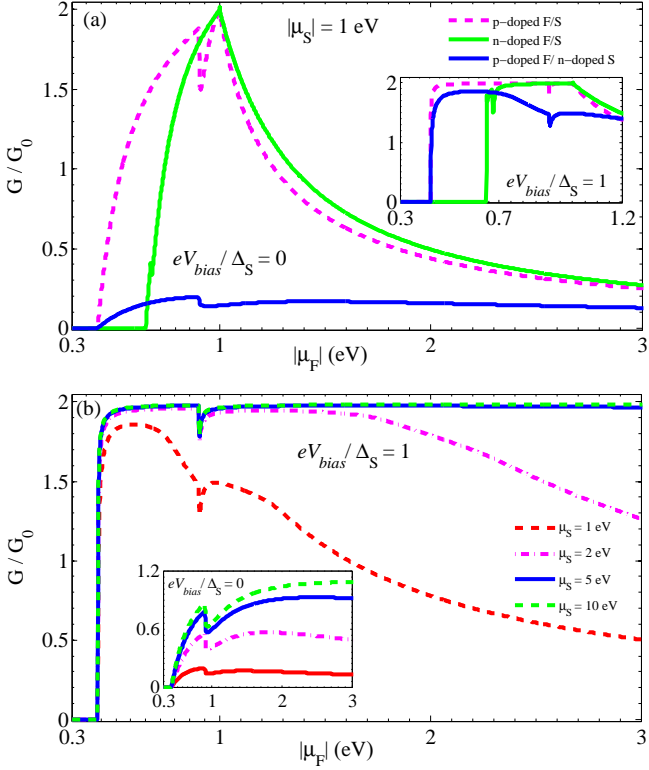}
\end{center}
\caption{\label{Fig:3} (Color online) (a) The zero-bias Andreev conductance versus the magnitude of the chemical potential inside the F region for three different unstrained WSe$_2$-based structures with $n$- or $p$-type doped F and S regions. Inset of (a) shows the corresponding results for $\textit{eV}_{bias}/\Delta_S=1$. (b) The behavior of the Andreev conductance for different values of the chemical potential $\mu_S$ in F/S structure with $n$-doped S and $p$-doped F regions, when $\textit{eV}_{bias}/\Delta_S = 1$. Inset of (b) shows the corresponding results for $\textit{eV}_{bias}/\Delta_S = 0$.}
\end{figure}
Turning on the SOC term in both F and S regions (WSe$_2$-based structure) causes the cusp-like behavior of the Andreev conductance as well as an Andreev peak at $|\mu_F|=|\mu_S|$ [see Fig. \ref{Fig:2}(b)]. Thereupon, the appearance of the cusplike behavior in the Andreev conductance is due to the presence of SOC in the F region. Note that in the absence of SOC, the valence-band edges for spin-$s$ and $\bar{s}$ subbands of both valleys lie at energies $E_v^{s,\tau(\bar{\tau})}=-B_v-\Delta/2$ and $E_v^{\bar{s},\tau(\bar{\tau})}=B_v-\Delta/2$, respectively. Since an electron-hole conversion via AR occurs for electrons and holes from opposite spin subbands with different valley indices [see Eq. (\ref{DBDG2})], there will be no possibility of AR process for incoming electrons from spin-$s$ subbands of $\tau$ and $\bar{\tau}$ valleys with the chemical potential in the range $E_v^{\bar{s},\tau(\bar{\tau})}-\textit{eV}_{bias}<\mu_F\leq E_v^{{s},\tau(\bar{\tau})}-\textit{eV}_{bias}$, while for $\mu_F\leq E_v^{\bar{s},\tau(\bar{\tau})}-\textit{eV}_{bias}$, the incident electron and the reflected hole can be from both spin subbands. Hence the AR process will be nonspin valley polarized and the resulting Andreev conductance will be increased monotonically with the chemical potential for $|\mu_F|\leq|\mu_S|$. However, as explained in Sec. \ref{sec:II}, a crossover from a spin valley polarized AR to that of a nonspin valley polarized one occurs with increasing the chemical potential $|\mu_F|$ in the presence of the spin-orbit interaction, that is responsible for a cusplike behavior of the Andreev conductance at the edge of the spin-$\bar{s}$ valley-$\tau$ of the valence band with $E_v^{\bar{s},\tau}=B_v-\lambda_v-\Delta/2$.
\par
\begin{figure}[]
\begin{center}
\includegraphics[width=3.4in]{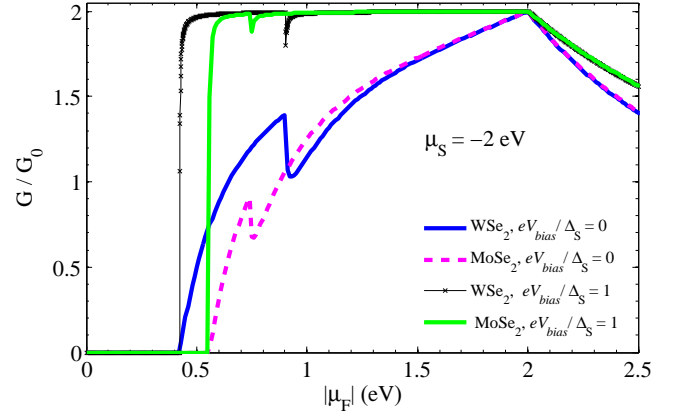}
\end{center}
\caption{\label{Fig:4} (Color online) Andreev conductance as a function of $|\mu_F|$ for $p$-doped WSe$_2$- and MoSe$_2$-based F/S structures with $\mu_S=-2$ eV when $\textit{eV}_{bias}/\Delta_S=0$ and 1.}
\end{figure}
Twisting the CrI$_3$ with $30^{\circ}$ relative to WSe$_2$ abandons the AR process for a wider range of $\mu_F$, shifts the cusp to larger $\mu_F$, attenuates the conductance of the spin valley polarized AR process, prevents perfect AR, and makes the Andreev conductance of the nonpolarized AR process (after the cusp) to be increased or decreased depending on the value of $\mu_F$ [see Fig. \ref{Fig:2}(c)]. Most importantly, as illustrated in Fig. \ref{Fig:2}(d), attenuation of the Andreev conductance in the presence of the twisting possesses its maximum value at zero bias $\textit{eV}_{bias}/\Delta_S=0$, decreases with enhancing the subgap bias and tends to zero for $\textit{eV}_{bias}/\Delta_S=1$. However, in the case of $\mu_F= -0.91$ eV (cusp position), twisting leads to an amplification of the Andreev conductance even at $\textit{eV}_{bias}/\Delta_S=1$.
\par
In addition, we present the effect of the type of charge doping on the Andreev conductance of a WSe$_2$-based F/S structure in Fig. \ref{Fig:3}(a), when $|\mu_S|=1$ eV and $\textit{eV}_{bias}/\Delta_S =0$ and $1$. In the case of $n$-type doping, a wide zero conductance gap, spin valley polarized AR process for a small range of $|\mu_F|$, and enhanced zero-bias Andreev conductance for $|\mu_F|>|\mu_S|$ are seen. Significantly, AR process with unit efficiency occurs for a smaller range of $|\mu_F|$ in contrast with that of the $p$-type structure when $\textit{eV}_{bias}/\Delta_S = 1$. On the other hand, a strong reduction of the zero-bias Andreev conductance, as well as different qualitative behavior with nonperfect AR for $\textit{eV}_{bias}/\Delta_S =1$, can be seen in the corresponding structure with $p$-doped F and $n$-doped S regions. Interestingly, we find that the amplification of the Andreev conductance and accordingly the AR process with unit efficiency (almost for all values of $\mu_F$) can be achieved for $p$-doped F/$n$-doped S junction by enhancing the magnitude of the chemical potential inside the S region [see Fig. \ref{Fig:3}(b)]. We have investigated the effects of twisting, type of charge doping and the gate electric field on the Andreev conductance of the corresponding MoSe$_2$-based F/S structure in Appendix \ref{sec:appendix A}. Importantly, the absence of the SOC term in the F region leads to a slight reduction of the Andreev conductance in contrast with that of the corresponding WSe$_2$-based structure. Astonishingly, a twist angle induced amplification of the Andreev conductance will be increased with the subgap bias for the chemical potential being at the cusp position. Comparing the results with those of the WSe$_2$-based structure demonstrates the suppression of the AR for a broad range of the chemical potential $\mu_F$ as well as the reduction of the Andreev conductance for other values of $\mu_F$ in the MoSe$_2$-based structure [see Fig. \ref{Fig:4}]. Note that due to the strong spin-orbit interaction in the  WSe$_2$, the spin-splitting of the valence band in WSe$_2$-based structure is considerably larger than that of the MoSe$_2$ and causes the cusplike behavior at larger values of $\mu_F$ compared with that of the MoSe$_2$-based structure.

\subsection{Thermal conductance}

\begin{figure}[]
\begin{center}
\includegraphics[width=3.5in]{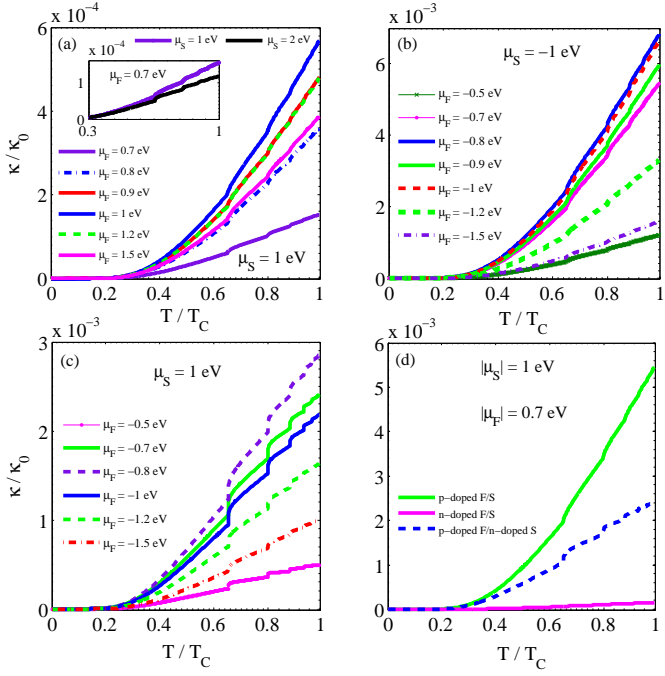}
\end{center}
\caption{\label{Fig:5} (Color online) The behavior of the normalized thermal conductance in terms of the temperature $T/T_C$ for different values of the chemical potential $\mu_F$ in untwisted WSe$_2$-based (a) $n$-doped F/S, (b) $p$-doped F/S, (c) $p$-doped F/$n$-doped S junction when $|\mu_S|=1$ eV. Inset of (a) shows the behavior of the thermal conductance versus $T/T_C$ for two values of $\mu_S$, when $\mu_F = 0.7$ eV. The comparison of the thermal conductance for different types of doping in F/S structure are illustrated in (d), when $|\mu_F|=0.7$ eV and $|\mu_S|=1$ eV.}
\end{figure}

\begin{figure}[]
\begin{center}
\includegraphics[width=3.5in]{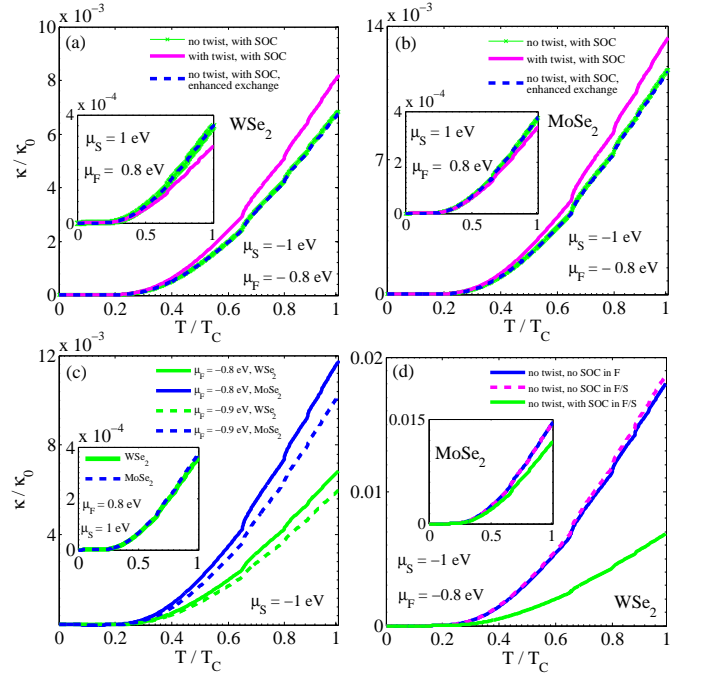}
\end{center}
\caption{\label{Fig:6} (Color online) The effects of the twist angle between the CrI$_3$ and the TMDC layer and the enhanced exchange coupling (of about 4 meV by positive gate electric field) on the thermal conductance of untwisted (a) WSe$_2$- and (b) MoSe$_2$-based $p$-doped F/S structures with $\mu_F=-0.8$ eV and $\mu_S=-1$ eV. Insets show the corresponding results for $n$-doped F/S junctions. The temperature dependence of the thermal conductance (c) for two values of $\mu_F$ in untwisted WSe$_2$- and MoSe$_2$-based $p$-doped F/S structures and (d) in the absence and presence of SOC in untwisted WSe$_2$-based $p$-doped F/S junction when $\mu_F=-0.8$ eV and $\mu_S=-1$ eV. Insets of (c) and (d), respectively, show the corresponding results for $n$-doped F/S junctions and for the MoSe$_2$-based F/S junction.}
\end{figure}

We now turn to investigate the thermal transport properties of the proposed F/S structure. Applying a temperature gradient $\Delta T$ through the junction, we can calculate the thermal conductance $\kappa=\lim_{\Delta T\rightarrow 0} J_{\textit{th}}/\Delta T $, with $J_{\textit{th}}$ the heat current density, as follows~\cite{Bardas,Yokoyama}
\begin{eqnarray}
\label{kappa}
\kappa&=&\frac{k_B W}{4\pi^2 \hbar}\sum_{s,\tau=\pm1}\int_{0}^{\infty}\int_{0}^{\pi/2}d\varepsilon\ d\theta_{s,\tau}^e \frac{\varepsilon^2\ |\bm{k}_{s,\tau}^e(\varepsilon)|\ \cos{\theta_{s,\tau}^e}}{(k_B T)^2{\cosh^2(\frac{\varepsilon}{2k_B T})}}\nonumber\\
&\times&[1-|r_{s,\tau}^e(\theta_{s,\tau}^e,\varepsilon)|^2-Re(\frac{\cos{\theta_{s,\tau}^h}}{\cos{\theta_{s,\tau}^e}})\ |r_{\bar{s},\bar{\tau}}^h(\theta_{s,\tau}^e,\varepsilon)|^2],\nonumber\\
\end{eqnarray}
where $T$ is the thermal equilibrium temperature and we replace the zero-temperature superconducting order parameter $\Delta_S$ in Eq. (\ref{DBDG2}) with the temperature-dependent one, $\Delta_S(T)=1.76\ k_B T_C\tanh{(1.74\sqrt{{T_C}/{T}-1})}$. Also, we set $k_B = 1$.

Figures \ref{Fig:5}(a)-\ref{Fig:5}(c) illustrate the behavior of the normalized thermal conductance $\kappa/\kappa_0$ [with $\kappa_0={k_B W}\sum_{s,\tau=\pm1}\int_{0}^{\infty}d\varepsilon |\bm{k}_{s,\tau}^e(\varepsilon)|/{4\pi^2 \hbar}$] in terms of the temperature $T/T_C$, respectively, for $n$-doped F/S, $p$-doped F/S, and $p$-doped F/$n$-doped S junctions with various values of the chemical potential $\mu_F$, when $|\mu_S|=1$ eV. It is noted that the thermal conductance, in contrast with the Andreev conductance, is suppressed by AR at low temperatures. The physical reason is that Cooper pairs carry a finite charge (2e) but zero heat across the junction. For the thermal conductance to be finite, the temperature must be so high that electronlike and holelike quasiparticles can be transmitted into the S region. Therefore the more significant transmission of quasiparticles at higher temperatures results in an increasing behavior of the thermal conductance with increasing the temperature. It turns out that increasing the magnitude of the chemical potential inside the F region leads to the enhancement of the thermal conductance of the $n$-doped F/S ($p$-doped F/$n$-doped S) structure for $\mu_F\leq\mu_S$ ($|\mu_F|\leq 0.8\ \mu_S$) and reduction of it for $\mu_F>\mu_S$ ($|\mu_F|> 0.8\ \mu_S$) [see Figs. \ref{Fig:5}(a) and \ref{Fig:5}(c)]. In contrast, varying the type of doping in F and S regions from $n$-type to $p$-type causes the thermal conductance of the $p$-doped F/S structure to be increased or decreased with enhancing the chemical potential $|\mu_F|$ [see Fig. \ref{Fig:5}(b)]. Moreover, it is singled out from the inset of Fig. \ref{Fig:5}(a) that enhancing the chemical potential $\mu_S$ tends to reduce the thermal conductance. Comparing the results of thermal conductance for F/S structures with various types of doping for F and S regions show that, in contrast with the Andreev conductance, the $n$-doped F/S structure has a very small thermal conductance and importantly the $p$-doped structure possesses a significant enhancement for the thermal conductance in comparison with that of the $p$-doped F/$n$-doped S structure [see Fig. \ref{Fig:5}(d)]. Similar results are obtained for the corresponding MoSe$_2$-based F/S junctions with $n$-type S region [see Appendix \ref{sec:appendix B}]. The difference is that enhancing the value of the chemical potential inside the S region tends to increase the thermal conductance. Also, the behavior of the thermal conductance of the $p$-doped F/S structure is similar to that of the $p$-doped F/$n$-doped S structure.

\begin{figure}[]
\begin{center}
\includegraphics[width=3.4in]{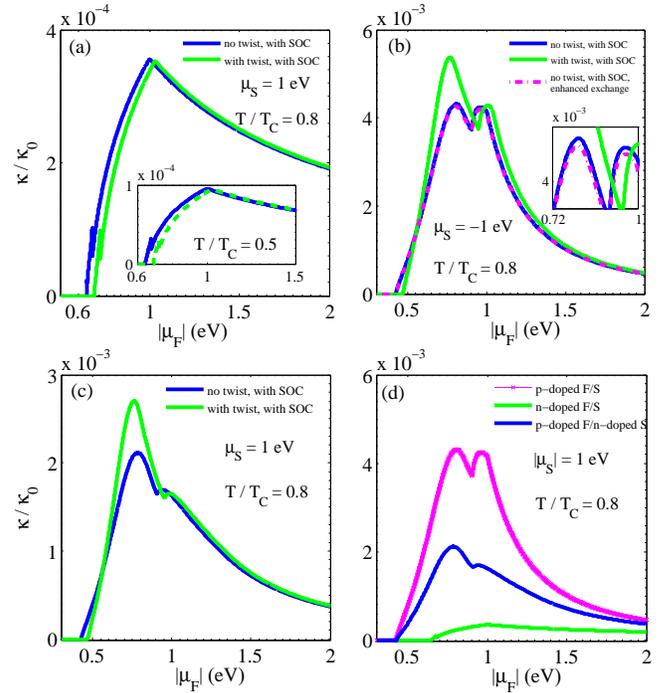}
\end{center}
\caption{\label{Fig:7} (Color online) The chemical potential dependence of the thermal conductance in the absence and presence of the twist angle between the CrI$_3$ and the WSe$_2$ layer in (a) $n$-doped F/S, (b) $p$-doped F/S, and (c) $p$-doped F/$n$-doped S structures with $|\mu_S|=1$ eV when $T/T_C = 0.8$. (d) Comparing the thermal conductance of F/S structures with various types of doping. Insets of (a) and (b), respectively, present the chemical potential dependence of the thermal conductance for $T/T_C = 0.5$ and the zoomed-in view of the $\kappa/\kappa_0$ ratio in the range 0.72 eV$\leq|\mu_F|\leq$1 eV.}
\end{figure}
We further evaluate the effect of the twist angle as well as the gate electric field on the thermal conductance of WSe$_2$- and MoSe$_2$-based F/S junctions, respectively, in Figs. \ref{Fig:6}(a) and \ref{Fig:6}(b). It is worth noting that twisting tends to enhance the thermal conductance of both WSe$_2$- and MoSe$_2$-based structures with $p$-type doping, whereas it reduces the thermal conductance of the corresponding structures with $n$-type doping. The magnitude of the proximity exchange coupling increases by applying a positive gate electric field and produces an extremely small reduction in the thermal conductance of both WSe$_2$- and MoSe$_2$-based structures with $p$-type doping. Comparing the results demonstrates that replacing the WSe$_2$ layer with MoSe$_2$ results in an enhancement of the thermal conductance of the $p$- or $n$-type doped structure [see Fig. \ref{Fig:6}(c)]. More importantly, it is shown in Fig. \ref{Fig:6}(d) that the thermal conductance in the absence of the SOC term in the F region (magnetized gapped graphene/ superconducting WSe$_2$ structure) as well as the F/S structure without SOC term (gapped graphene-based structure) are significantly larger than that of the WSe$_2$-based structure [see Fig. \ref{Fig:6}(d)]. Meanwhile, the inset of Fig. \ref{Fig:6}(d) presents less enhancement of the thermal conductance in the corresponding structures based on MoSe$_2$.

Ultimately, we present the chemical potential dependence of the thermal conductance of WSe$_2$-based F/S structures with various types of doping for the F and S regions in Fig. \ref{Fig:7}, when $|\mu_S| = 1$ eV. In the case of the $n$-doped F/S structure, it is perceived that the thermal conductance experiences a cusplike behavior, increases with $\mu_F$ for $\mu_F<\mu_S$ and after reaching a maximum at $\mu_F=\mu_S$ it decreases for $\mu_F>\mu_S$ [ see Fig. \ref{Fig:7}(a)]. Twisting the CrI$_3$ with respect to the WSe$_2$ layer suppresses the thermal conductance for a broader range of $\mu_F$. Increasing the chemical potential of the F region makes the thermal conductance decrease relative to the untwisted one for $\mu_F<\mu_S$. The twist-angle induced changes in the magnitude of the thermal conductance decrease with increasing $\mu_F$ and tend to zero at low $T/T_C$ ratio, when $\mu_F\rightarrow\mu_S$ [see the inset of Fig. \ref{Fig:7}(a)]. Nevertheless, increasing the $T/T_C$ ratio can lead to an enhancement of the thermal conductance in the presence of the twisting for $\mu_F>\mu_S$. Figures \ref{Fig:7}(b) and \ref{Fig:7}(c) present different peak structures for the thermal conductance of F/S structures with the $p$-doped F region and indicate that twisting can lead to significant amplification of the thermal conductance for the chemical potentials $|\mu_F|<|\mu_S|$. Moreover, it is noted from the inset of Fig. \ref{Fig:7}(b) that the gate electric field-induced enhancement of the proximity exchange coupling results in a slight reduction of the thermal conductance. Comparing the results of thermal conductance for various types of doping in F/S structures confirms that the high thermal conductance belongs to $p$-doped structure and the F/S structure with the $n$-type doping has low thermal conductance [see Fig. \ref{Fig:7}(d)].

\section{Conclusion}\label{sec:IV}

In summary, we have studied the electrical and thermal transport properties of a TMDC/CrI$_3$ heterostructure (as ferromagnetic region) connected to a TMDC superconducting layer within the scattering formalism. It is found that the presence of a spin-orbit interaction leads to a spin valley polarized AR process, which can be effectively modulated by tuning the charge doping, twisting the CrI$_3$ monolayer relative to the TMDC layer and applying a gate electric field. The perfect spin valley polarized AR is visible for subgap biases $\textit{eV}_{bias}<\Delta_S$ when the chemical potential of the ferromagnetic domain $\mu_F$ in a $p$- or $n$-doped structure is the same as that of the superconducting region $\mu_S$ and over a wide range of $\mu_F$ ($|\mu_F|\leq|\mu_S|$) when $\textit{eV}_{bias}=\Delta_S$. Also, it can be achieved for almost all values of $\mu_F$ in the corresponding structure with different types of charge doping for F and S regions by increasing the chemical potential of the S region, if $\textit{eV}_{bias} =\Delta_S$.

Importantly, the twist angle between the CrI$_3$ and the TMDC (with WSe$_2$ and MoSe$_2$ as TMDC) leads to the suppression or reduction of the spin valley polarized Andreev conductance and the enhancement of the nonspin valley-polarized Andreev conductance, depending on the chemical potential $\mu_F$. Notably, changes in the magnitude of the Andreev conductance by twisting are reduced to an increase in the subgap bias and might vanish at $\textit{eV}_{bias}=\Delta_S$ in the WSe$_2$-based structure. However, it can be enhanced by increasing the bias for the defined chemical potential of the ferromagnetic region in the MoSe$_2$-based structure. We further analyzed the influence of twisting on thermal conductance and showed that for $|\mu_F|<|\mu_S|$ in the proposed structure with $p$-type doping, an increase in thermal conductance can occur while it is reduced in the $n$-doped structure. Furthermore, increasing the proximity exchange by applying a positive gate electric field to the TMDC/CrI$_3$ heterostructure slightly decreases the thermal conductance of the $p$-type doping as well as the probability of an AR process and accordingly the Andreev conductance and suppresses them for small values of $\mu_F$.

Furthermore, we have shown that Andreev and thermal conductance can be significantly altered by tuning the type of charge doping in ferromagnetic and superconducting regions. The proposed structure shows high Andreev and thermal conductance in the case of $p$-type doping. We have further found that the low charge conductance of the $p$-doped ferromagnetic/$n$-doped superconducting junction can be increased remarkably by increasing the chemical potential of the superconducting region. Furthermore, our results show that replacing WSe$_2$ with MoSe$_2$ tends to increase the Andreev conductance as well as enhancing thermal conductance in $p$-doped F/S structures and decreasing thermal conductance in $n$-type doped structures. In addition, changing the TMDC material leads to different dependencies of the chemical potential $\mu_F$ of the thermal conductance. Furthermore, switching off the spin-orbit interaction in the structure by using graphene with voids in one or both of the ferromagnetic and superconducting regions causes significant changes in the electrical and thermal conductance of the proposed structure; it leads to a significant damping of the Andreev conductance, as well as a large increase in thermal conductance. Our theoretical finding can be explored through actual experiments.

\section*{Acknowledgments}
This work is supported by Iran Science Elites Federation (Grant No. M1400138).

\appendix

\section {\label{sec:appendix A} The effects of spin-orbit interaction, twisting and the applied transverse-electric field on the Andreev conductance of a MoSe$_2$-based F/S structure}
\begin{figure}[]
\begin{center}
\includegraphics[width=3.4in]{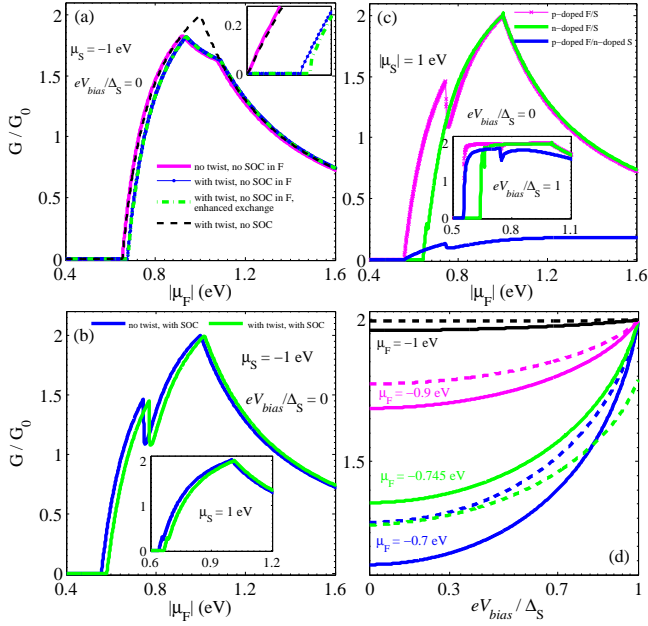}
\end{center}
\caption{\label{Fig:8} (Color online) (a) The zero-bias Andreev conductance as a function of the chemical potential $|\mu_F|$ in the (a) absence and (b) presence of the SOC inside the F region of the MoSe$_2$-based $p$-doped F/S structure when $\mu_S=-1$ eV. Enhancement of the magnitude of the proximity exchange interaction in the presence of the positive transverse-electric field is about $4$ meV. (c) The zero-bias Andreev conductance versus $|\mu_F|$ for three different unstrained MoSe$_2$-based structures with $n$- or $p$-type doped F and S regions. Inset shows the corresponding results for $\textit{eV}_{bias}/\Delta_S=1$. (d) The bias voltage dependence of the Andreev conductance for different values of the chemical potential $\mu_F$ in untwisted (dashed lines) and twisted (solid lines) $p$-doped MoSe$_2$-based structures when $\mu_S = -1$ eV.}
\end{figure}
We investigate the behavior of the Andreev conductance of a $p$-doped MoSe$_2$-based F/S structure in Fig. \ref{Fig:8}. In comparison with the WSe$_2$-based structure, a slight reduction of the Andreev conductance in the absence of the SOC term in the F region is obvious from Fig. \ref{Fig:8}(a). A cusplike behavior is present away from the Andreev peak, at a minor value of the chemical potential inside the F region, because of the weaker spin-orbit interaction in MoSe$_2$ [see Fig. \ref{Fig:8}(b)]. Importantly, an amplification of the Andreev conductance by twisting increases with the subgap bias for the chemical potential $\mu_F=-0.745$ eV, being at the cusp position, while it decreases with $\textit{eV}_{bias}/\Delta_S$ for WSe$_2$-based structure [see Fig. \ref{Fig:8}(d)]. In addition, it can be perceived from Fig. \ref{Fig:8}(c) that the $p$-doped F/S junction has higher spin valley polarized Andreev conductance in comparison with the $n$-type one at zero bias and perfect electron-hole conversion occurs for a wide range of $\mu_F$, when $\textit{eV}_{bias}/\Delta_S = 1$.

\section {\label{sec:appendix B} Thermal conductance of MoSe$_2$-based F/S structures with various types of charge doping for F and S regions}
\begin{figure}[]
\includegraphics[width=3.5in]{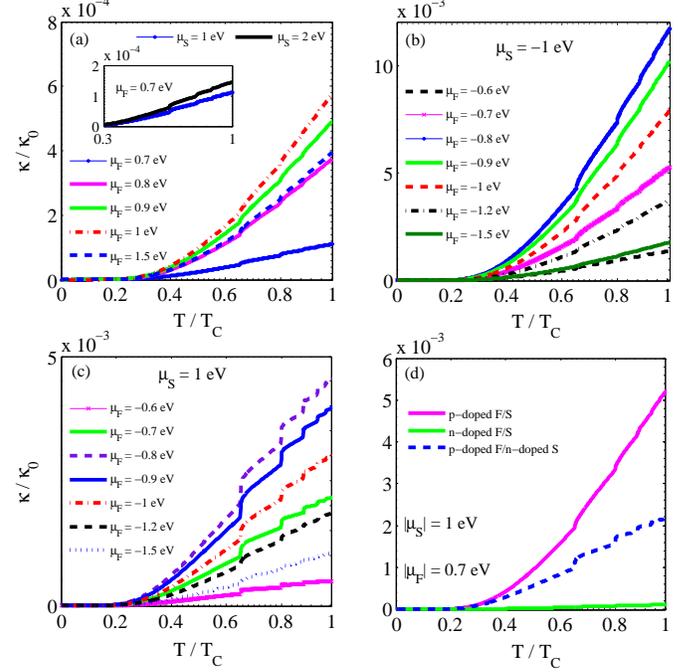}
\caption{\label{Fig:9}(Color online) Thermal conductance in terms of the temperature $T/T_C$ for different values of the chemical potential $\mu_F$ in an untwisted MoSe$_2$-based (a) $n$-doped F/S, (b) $p$-doped F/S, and (c) $p$-doped F/$n$-doped S junction, when $|\mu_S|=1$ eV. (d) Comparison of the thermal conductance for F/S structures with different types of doping, when $|\mu_F|=0.7$ eV and $|\mu_S|=1$ eV.}
\end{figure}

The behavior of the normalized thermal conductance $\kappa/\kappa_0$ in terms of the temperature $T/T_C$ are illustrated in Figs. \ref{Fig:9}(a)-\ref{Fig:9}(c),   respectively, for the MoSe$_2$-based $n$-doped F/S, $p$-doped F/S, and $p$-doped F/$n$-doped S junctions with various values of the chemical potential $\mu_F$, when $|\mu_S|=1$ eV. The thermal conductance has an increasing behavior with the temperature. Similar to the WSe$_2$-based structures, the thermal conductance increases with enhancing the magnitude of the chemical potential inside the F region of $n$-doped F/S ($p$-doped F/$n$-doped S) structure for $\mu_F\leq\mu_S$ ($|\mu_F|\leq 0.8\ \mu_S$) and decreases for $\mu_F>\mu_S$ ($|\mu_F|> 0.8\ \mu_S$). However, in contrast with the WSe$_2$-based structure, the thermal conductance of the $p$-type doped structure increases with enhancing $|\mu_F|$ for $|\mu_F|\leq 0.8\ |\mu_S|$ and decreases for $|\mu_F|> 0.8\ |\mu_S|$. Besides, the thermal conductance can be increased by enhancing the chemical potential of the S region [see the inset of Fig. \ref{Fig:9}(a)]. We further demonstrate the amplification of the thermal conductance in F/S structures with a $p$-doped F region in comparison with that of the $n$-type doping [see Fig. \ref{Fig:9}(d)].

\begin{figure}[]
\includegraphics[width=3.5in]{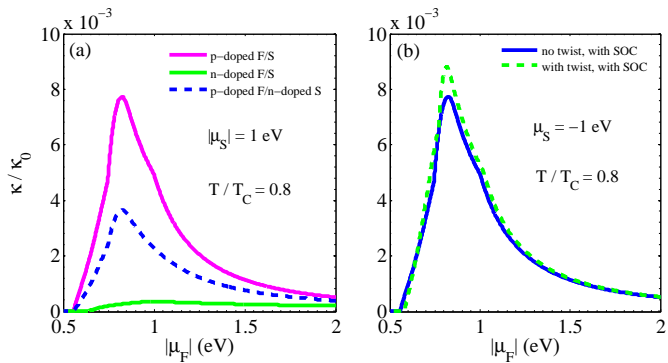}
\caption{\label{Fig:10}(Color online) The chemical potential dependence of the thermal conductance of MoSe$_2$-based F/S structures (a) with various types of doping
for F and S regions and (b) in the absence and presence of twisting for $p$-type doping when $|\mu_S|=1$ eV and $T/T_C=0.8$.}
\end{figure}

Interestingly, as it is presented in Fig. \ref{Fig:10}, the peak structures of the thermal conductance in terms of $|\mu_F|$ in MoSe$_2$-based F/S structures with various types of charge doping for F and S regions are different from those of the corresponding WSe$_2$-based structures. Also, the high thermal conductance of $p$-doped structure in comparison with that of the $n$-doped structure is obvious from Fig. \ref{Fig:10}(a). Moreover, it can be seen from Fig. \ref{Fig:10}(b) that twisting tends to the amplification of the thermal conductance in a wide range of $\mu_F$. However, we have found (not shown) that the enhancement of the proximity exchange by the positive gate electric field leads to the small reduction of the thermal conductance around the peak.

\end{document}